# Magnetic Phase transition and Relaxation effects in Lithium Iron Phosphate

Y. Sundarayya[1], C. Bansal[1], C.S. Sunandana[1]

[1] School of Physics, University of Hyderabad, Hyderabad – 500046, India.

Ajay Kumar Mishra[2], Richard A. Brand[2], Horst Hahn[2]

[2] Karlsruhe Institute of Technology (KIT), Institute of Nanotechnology, Karlsruhe, Germany.

**Abstract**

We report the observation of para – antiferromagnetic transition at ~ 50 K in lithium iron phosphate, LiFePO$_4$ through DC magnetization and Mössbauer spectroscopy. The Ferrous ion Fe$^{2+}$ $\left(3d^6, {}^5D\right)$ in LiFePO$_4$ exhibits relaxation effects with a relaxation frequency ~ 1.076 × 10$^7$ s$^{-1}$ at 300 K. The temperature dependence of the frequency suggests the origin of the relaxation is spin-lattice type. The quadrupole splitting at low temperatures indicates the excited orbital states mix strongly to the orbital doublet ground state via spin-orbit coupling. Modified molecular field model analysis yields a saturation value for hyperfine field $B_{hf}$ ~ 125 kOe. The anomaly in magnetization and Mössbauer parameters below 27 K may be ascribed to contribution of orbital angular momentum. The high value of the asymmetry parameter $\eta$ (~ 0.8) of the electric field gradient obtained in the antiferromagnetic regime indicates a strongly distorted octahedral oxygen neighbourhood for the ferrous sites.

1.  **Introduction**

Lithium iron Phosphate, LiFePO$_4$ is one among the general class of polyanion olivine type compounds containing tetrahedral phosphate structural unit that gives the structural stability with strong covalent bonding, networked to produce oxygen octahedra around Fe(+2) ion [1]. Ever since LiFePO$_4$ was found to be electrochemically active material, the research activity on this material has increased enormously. While it crystallizes in to orthorhombic phase with space group *Pnma*, the magnetic structure was found to undergo a transition from paramagnetic to antiferromagnetic at 50±2 K, the Néel temperature, $T_N$, with the spins oriented along the *b* crystallographic direction [2]. It was shown the magnetic properties of materials LiMPO$_4$ (M = Mn, Fe, Co, Ni) do exhibit similar magnetic phase transitions, during which they only differ in the orientation of the staggered spins [3 – 13]. On the other hand, it was claimed that the M–O–M superexchange interactions in the LiMPO$_4$ (M = Mn, Fe, Co, Ni) are responsible for the presence of antiferromagnetic puckered planes orthogonal to *a* [2]. Structurally LiFePO$_4$ has closely packed oxygen framework in which the ferrous ion resides in the form of distorted FeO$_6$ octahedron by sharing the oxygen with PO$_4$ tetrahedron. The *Fe$^{2+}$* ions (S = 2) form corrugated layers that are stacked along the [100] crystallographic axis. Nearest neighbors in the b-c plane are coupled magnetically by a relatively strong exchange interaction, $J_1$ through a Fe-O-Fe oxygen bond, where as in-plane next-nearest-neighbors are coupled ($J_2$) via Fe-O-O-Fe bonds [14]. Interlayer magnetic coupling is mediated by a phosphate ion via Fe-O-P-O-Fe bonding [15]. Thus the olivine family of LiMPO$_4$ exhibits highly anisotropic properties which are intermediate between those of two-(2D) and three-dimensional (3D) systems, LiFePO$_4$, in particular [3, 4].

The origin of magnetism in LiFePO$_4$ must be traced to the $Fe^{2+}$ moments stabilized in the orthorhombic lattice and the highly temperature sensitive microscopic crystalline environments in which they exist **[16]**. It is known that the magnetic properties are determined by the electronic states and may thus reflect the potential advantage of LiFePO$_4$. $^{57}$Fe Mössbauer spectroscopy is a powerful tool to investigate the dynamics associated with magnetic transitions. It provides information on the electronic level scheme of $Fe^{2+}$ in the crystal field approach and allows the electronic ground state to be determined at low-temperature **[17]**. The time scale $\tau_M$ of Mössbauer transitions is on the order of nanoseconds and therefore it allows determining a range of relaxation times that are not easy to measure by other conventional techniques. By the Mössbauer method the electronic spin relaxation frequency is estimated through the line-broadening of the absorption peaks **[18, 19]**. While the previous reports on the Mössbauer studies on LiFePO$_4$ unambiguously confirms the Para – antiferromagnetic transition around 50 K **[20 - 23]**, we discuss here the relaxation phenomenon associated with the $Fe^{2+}$ ions in LiFePO$_4$ that appear in the Mössbauer spectra recorded from 20 – 300 K through the magnetic transition.

2. **Experimental**

Polycrystalline sample of LiFePO$_4$ was synthesized by non-aqueous sol-gel route using reagent grade lithium acetate dihydrate (Loba chemicals), ferrous oxalate dihydrate (Riedel-de Haën), and ammonium dihydrogen phosphate (Merck) as precursors and the details are described elsewhere **[24]**. Phase purity was characterized by X-ray data collected with INEL X-ray diffractometer using $Co-K_\alpha$ radiation (λ = 1.78897 Å). Rietveld refinements were carried out from the X-ray data using a software package Fullprof **[25]**. The morphology of the samples was studied by means of field-emission scanning electron microscopy (FESEM, FEI NOVA NANOSEM 600). Room temperature FTIR spectrum was recorded on a NEXUS

FTIR spectrometer from Thermo–Nicolet with an extended range KBr (XT – KBr) beam splitter. DC magnetization measurements were carried out using a Superconducting Quantum Interference Device (SQUID) magnetometer under zero-field-cooled (ZFC) and field-cooled (FC) condition in the temperature range 5 – 300 K under a magnetic field of 100 Oe. 'Zero field' Mössbauer spectra were recorded in transmission mode using $^{57}$Co γ-ray source in a Rhodium matrix and multi-channel analyzer. The sample thickness was adjusted so that the Fe content was approximately 10 mg/cm$^2$. Measurements at low temperatures were carried out using a variable temperature helium cryostat insert, and the temperature was stabilised to ± 0.1 K.

## 3. Results and Discussion

Figure 1(a) shows the indexing of room temperature powder X-ray data of the sample suggested the possible orthorhombic phase of tryphilite $LiFePO_4$ with space group *Pnma* **[26]**. Table I summarises the structural parameters of $LiFePO_4$ at 300 K determined from Rietveld refinement on powder X-ray diffraction data. A detailed crystallite-size analysis of the X-ray diffracted pattern using Scherrer's formula **[27]** reveals that crystallites of $LiFePO_4$ with average sizes 45(5) nm are present. The atomic arrangement at different sites in the crystal structure of $LiFePO_4$ obtained from Rietveld refinement is depicted in the inset of Figure 1(a). The room temperature FTIR spectrum shown in Figure 1(b) depicts the bending and stretching modes of the tetrahedral phosphate group that networks to the $FeO_6$ octahedra to give $LiFePO_4$. The inset of Figure 1(b) shows the morphological image of $LiFePO_4$, indicates particles with size distribution of ~ 150 nm.

The field cooled (FC) DC magnetization measurements on $LiFePO_4$ is shown in Figure 2 in the temperature range 5 K < $T_N$ << 300 K. With decrease in temperature, the magnetization (M) increases down to 50K (= $T_N$) and suddenly decreases indicating the magnetic transition.

The inset of Figure 2 depicts the reciprocal magnetic susceptibility $\chi^{-1}(T)$ below $T_N$ and found $\chi^{-1}(T)$ undergoes an increase below 50 K [2, 20]. The magnetic susceptibility shows a deviation from the linear behavior at this temperature suggesting that the sample is well crystallized and neither structural defects nor the impurities can impede the propagation of long-range spin correlations that are responsible for the onset of the antiferromagnetic ordering. And also this spin correlation length is limited by thermal fluctuations. We note that the temperature dependence of the magnetic susceptibility does not follow the dependence predicted for the 2D square lattice of classical spins as it does not show a typical broad maximum [28]. The reciprocal magnetic susceptibility $\chi^{-1}(T)$ below $T_N$ is fitted to a straight line to obtain $\chi(T)$ at absolute zero and was found to be $7.937 \times 10^{-4}$ emu/gm Oe.

'Zero field' Mössbauer spectra of LiFePO$_4$ in the paramagnetic regime are shown in Figure 3(a). The spectra above the Néel temperature, typically at 300, 100, 55 and 50 K could be fitted with a Lorentzian-shaped doublet with reasonably narrow line widths. In the figure, the open circles show the data and the fits by thick line passing through the data. The Mössbauer parameters isomer shift $\delta$ and quadrupole splitting $\Delta E_0$ of the doublet at 300 K are found to be $\delta = 1.222$ mm/s and $\Delta E_0 = 3.000$ mm/s respectively. These values are typical for $Fe^{2+}$ in ferrous compounds as they exhibit lower electron density due to six $3d$ electrons on high spin $Fe^{2+}$. The asymmetric nature of the doublet at 300 K can be ascribed to the $Fe^{2+}$ in distorted FeO$_6$ octahedron [24, 29 and 30].

The spectra below Néel temperature are analysed by the diagonalization of the full static Hamiltonian, which includes both the magnetic dipole and electric quadrupole interaction. The Hamiltonian for the I = 3/2 first excited state of $^{57}$Fe may be written as [28, 31]

$$H = g_1 \mu_N I.H + \frac{e^2 qQ}{4I(2I-1)}[3I_{z'}^2 - I^2 + \eta(I_{x'}^2 - I_{y'}^2)] \qquad (1)$$

Where $x'$, $y'$ and $z'$ represent the principal axes of the electric field gradient (EFG) tensor, which are chosen so that $|V_{z'z'}| > |V_{y'y'}| \geq |V_{x'x'}|$, and where H is the internal magnetic field at the nucleus. The EFG along $z'$ is $eq$, and $\eta = |V_{x'x'} - V_{y'y'}|/V_{z'z'}$ is the asymmetry parameter. The $g_1$ is the g factor for the first excited state of $^{57}$Fe, $\mu_N$ is the nuclear magneton. $Q$ is the nuclear quadrupole moment of the first excited state of $^{57}$Fe multiplied by Sternheimer factor. This static full Hamiltonian properly calculates the line positions and intensities for any combination of hyperfine magnetic field $B_{hf}$ and electric field gradient (EFG) tensor, with $\Omega$ being the polar angle between the magnetic hyperfine field $B_{hf}$ and the direction of the principal component of electric field gradient.

Let us first look at the electronic structure of $Fe^{2+}(3d^6, {}^5D)$ with octahedral crystal field environment in $LiFePO_4$. The free-ion $(3d)^6, {}^5D$ state ($L = 2$, $S = 2$) of the $Fe^{2+}$ is split by the octahedral crystal field into the orbital doublet ($E$) and orbital triplet ($T2$), the latter being the lowest one for axial symmetry **[32, 33]**. By taking the easy axis (Ising) of $LiFePO_4$ as the quantization axis, the orbital wave function of the ground orbital triplet may be given by

$$|-1\rangle = |\Psi_{d1}\rangle = \sqrt{\frac{2}{3}}|\phi_{-2}\rangle + \frac{1}{\sqrt{3}}|\phi_{1}\rangle$$

$$|0\rangle = |\Psi_{s}\rangle = |\phi_{0}\rangle \quad (2)$$

$$|1\rangle = |\Psi_{d2}\rangle = \sqrt{\frac{2}{3}}|\phi_{2}\rangle - \frac{1}{\sqrt{3}}|\phi_{-1}\rangle$$

where we use the notation, $|l = 3, m\rangle = |\phi_m\rangle$ and we have the relation $\hat{L}_z|\phi_m\rangle = m|\phi_m\rangle$ (m = -3, -2, -1, 0, 1, 2, 3). This triplet ($T2$) splits into a doublet $|\Psi_{d1}\rangle$, $|\Psi_{d2}\rangle$ and a singlet $|\Psi_s\rangle$ in an octahedral crystal field.

For a well isolated orbital ground state belonging to $\Gamma_5$ manifold in axial symmetry, the principal component of the electric field gradient $V_{z'z'}$ (= $V_0$) takes the form $V_0 = -(4/7)e\langle r_Q^{-3}\rangle$ for orbital singlet $|\Psi_s\rangle$ and $V_0/2 = (2/7)e\langle r_Q^{-3}\rangle$ for orbital doublet $|\Psi_{d1}\rangle$, $|\Psi_{d2}\rangle$. This results a quadrupole interaction $\Delta E_Q$ amounting to a high value, decided by the ionic/ covalent nature of the environment around ferrous ion **[16]**.

When the orbital splitting is large enough, the spin-orbit coupling splits the ground orbital singlet into five degenerate spin levels and the ground-orbital doublet splits into five spin-orbit doublets. Here, the secondary effects from the singlet state may be neglected when the energy separation between the doublet and singlet is much higher than the spin-orbit coupling energy. Thus, the spin-orbital coupling which splits the ground-orbital doublet into five spin-orbit doublets is given by $H_0 = -k\lambda \vec{l}\cdot\vec{S}$. Here k ($\approx$ 1) is constant, $\lambda$ ($\sim$ $-100$ cm$^{-1}$ for Fe$^{2+}$) is a spin-orbit coupling constant, and $S$ (=2) is the spin angular momentum. The lowest of these doublets has the wave functions $\Psi_{d1}\chi_{-2}$ and $\Psi_{d2}\chi_{-1}$. The first excited doublet may be expressed as a linear combination of ground doublet and has the wave functions $\varphi_1 = (\Psi_{d1}\chi_{-1} - \Psi_{d2}\chi_1)/\sqrt{2}$ and $\varphi_2 = (\Psi_{d1}\chi_{-1} + \Psi_{d2}\chi_1)/\sqrt{2}$, where $\chi_m$ ($m$ = 0, ±1, ±2) are the spin wave functions of $S$ = 2. The excited doublet has a separation $\lambda$ with ground doublet. The temperature dependence of the quadrupole splitting is interpreted within the $^5$D orbital energy level scheme of $Fe^{2+}$ by a crystal field calculation based on the point symmetry $C_s$ of the $Fe^{2+}$ site in the LiFePO$_4$ have been discussed elsewhere **[21]**.

The results obtained from the Mössbauer spectral analysis are summarized as follows. In the paramagnetic regime 50 – 300 K, the quadrupole splitting $\Delta E_Q$ of LiFePO$_4$ $\sim$ 3.000 mm/s, is nearly constant and positive that the doublet could give. This indicates that the energy separation between the doublet ($|\Psi_{d1}\rangle$, $|\Psi_{d2}\rangle$) and the singlet ($|\Psi_s\rangle$) is of the order of spin-orbit coupling energy and excited orbital states mix strongly with the ground state via spin-orbit coupling. A decrease of $\Delta E_0$ to a value ~ 2.785 mm/s is observed in the antiferromagnetic regime below 50 K confirms the orbital doublet is lower than the singlet, typical for axial symmetry of the high-spin ferrous ion in the ground state [33]. Figure 4(a) shows the variation of the quadrupole splitting with temperature. The temperature dependence of the quadrupole splitting is due to thermal averaging over the crystalline field level of the $Fe^{2+}$ ion, the perturbation of the ion by magnetic exchange and spin-orbit coupling, and the EFG parameters due to the lattice [22].

At 300 K, the asymmetric nature of the doublet may be ascribed to the $Fe^{2+}$ in distorted $FeO_6$ octahedron. Further, higher velocity line that corresponds to $3/2 \rightarrow 1/2$ transition undergoes asymmetric broadening with decreasing temperature. The degree of asymmetry, though small, increases in the paramagnetic region upon approaching towards $T_N$ (~ 50 K), as shown in Figure 3(a). The origin of the asymmetry of the doublet components may be attributed to slow electron-spin relaxation that occurs when the electron spin flips at a rate comparable to the nuclear precessional frequency [34]. The different Mössbauer parameters thus obtained in the paramagnetic regime are given in table II. It is noteworthy to mention here that the observed asymmetric broadening is opposite to *Goldanskii–Karyagin* effect, for which a decrease of asymmetry is observed with decreasing temperature [35].

The 'Zero field' Mössbauer spectra below Néel temperature could be fitted with well-resolved octuplet, typical for magnetically ordered single-phase ferrous ion in $LiFePO_4$. The spectra were analyzed by the diagonalization of the $4 \times 4$ magnetic hyperfine and quadrupole-interaction Hamiltonian matrix for the first excited state of $^{57}Fe$ to give magnetic hyperfine field and quadrupole splitting, as given in eq. (1). The Mössbauer spectra of $LiFePO_4$ in the antiferromagnetic regime, typically for temperatures 20, 35, 45 and 48 K with the respective Hamiltonian fits are shown in Figure 3(b). The line positions and relative intensities are calculated for $\Omega = 0^o$ i.e., the hyperfine field $B_{hf}$ is along the principal axis of the electric field gradient. This is due to the fact that the EFG axis must be perpendicular to the mirror symmetry plane, so are the magnetic moments. This conclusion is based on the neutron diffraction and other Mössbauer analysis reports which infer that the ferrous magnetic moments in $LiFePO_4$ are aligned along the *b*-axis. Otherwise there are families of solutions involving hyperfine field, quadrupole splitting and polar angle all of which give strictly the same Mössbauer pattern at a given temperature. The different Mössbauer parameters thus obtained in the antiferromagnetic regime are given in table III. It is to be noted for $Fe^{3+}$ ion with axially symmetric EFG and weak quadrupole interaction results in a pure sextet.

Adopting the above iteration procedure to the low temperature Mössbauer patterns results in fits with adjusted line shapes and intensities that clearly show slight deviation from the observed patterns. This could be due to the constraint imposed on the fit by considering one single component with constant line width for all the eight transitions and/or from the feature that one or more hyperfine parameters to some extent exhibits a distribution. Van Albloom et al have considered that these misfits could have arisen due to structural imperfections, locally varying bonding properties or stoichiometric deviations due to which, the ferrous hyperfine parameters becomes sensitive leading to misfits [21]. To ascertain whether or not the deviation of the fit arises from the sample alone, a paramagnetic doublet corresponding with ferrous Mössbauer parameters was fitted along with the Hamiltonian matrix. It was found the final fit also shows a deviation from the observed, though the *chi²* of the fit slightly reduced. In addition to this, the misfit parts obtained by subtracting the theoretical spectra thus generated from the experimental spectra show that the component spectra corresponding to ±1/2 electronic states are highly affected by relaxation even when the relaxation times of ±5/2 and ±3/2 states of the $Fe^{2+}$ ions are large. An observable asymmetric broadening of the highest energy line is clearly shown in the Mössbauer patterns below Néel temperature. It is important to note that this observation is not due to instrumental and/ or non uniform temperature distribution across the sample. The high value of the asymmetry parameter $\eta$ (~ 0.8) of the EFG obtained in the antiferromagnetic regime indicates strongly distorted octahedral oxygen for the ferrous sites. It is to be mentioned that the lamb-Mössbauer parameter $f_{300K}$ (= 0.73) and Debye temperature $\theta_M$ (= 336 K) for LiFePO$_4$ are estimated from low temperature Mössbauer spectra elsewhere [36].

The temperature dependence of the hyperfine field $B_{hf}$ for LiFePO$_4$ is shown in Figure 4(b). It seems to obey molecular field theoretical curve in general. The precise dependence of mean $B_{hf}$ is analyzed by the following relation [37]

$$B_{hf} = B_{hf}(T=0)\left\{1-\left(\frac{T}{T_N}\right)^p\right\}^q \qquad (3)$$

where $p$ and $q$ are fitting parameters, $T_N$ is Néel temperature and $B_{hf}$ is hyperfine field. In Figure 4(b), the green and red lines show the fit and the data. The fitting parameters obtained from the above fit are $p = 16.43$, $q = 2.1$, $T_N$ ~ 50 K. The saturation value for hyperfine field at absolute zero is found to be $B_{hf}$ (T = 0 K) ~ 125 kOe and is in good agreement with the literature [21].

Due to the non-spherical nature, ferrous ions are strongly coupled to lattice vibrations resulting in spin-lattice relaxation (SLR) that may be observed by the asymmetric broadening of the Mössbauer lines [19]. The width of the absorption line on the higher velocity side is used to estimate the relaxation frequency $\left(\tau^{-1}\right)$ of the Mössbauer spectra of LiFePO$_4$ at 300 K down to 20 K. In the paramagnetic regime, the relaxation frequency at 300 K is obtained to be $1.076 \times 10^7$ s$^{-1}$ and found to decrease to $1.032 \times 10^7$ s$^{-1}$ at 50 K. This spin-lattice relaxation of high-spin ferrous ion at ambient temperature is rapid and due to temperature dependence of the population of the phonons, the spin-lattice relaxation frequency $\left(\tau_{sl}^{-1}\right)$ becomes slower at lower temperatures. In the antiferromagnetic regime, the relaxation frequency at 45 K is obtained to be $1.029 \times 10^7$ sec$^{-1}$ and found to increase to $1.199 \times 10^7$ s$^{-1}$ at 20 K. The temperature dependence of relaxation frequency $(\tau^{-1})$ below and above Néel temperature $T_N$ is shown Figures 4(c) and 4(d). This demonstrates the relaxation frequency obeys the spin-lattice type even at temperatures down to 35 K.

Finally, an anomalous drop in $\chi^{-1}(T)$ is observed for LiFePO$_4$ at T ~ 27 K upon cooling through this temperature. This feature signals the onset of a spontaneous magnetization at this temperature [10]. This is ascribed to the presence of weak ferromagnetic component of Li$_3$Fe$_2$(PO$_4$)$_3$ by showing a narrow doublet in the room temperature Mössbauer spectrum [28, 38]. This narrow doublet is conspicuously absent in the room temperature Mössbauer spectrum of LiFePO$_4$ rules out the presence of Li$_3$Fe$_2$(PO$_4$)$_3$ in our sample. However, we observed a sudden increase of $\tau^{-1}$ below 35 K, corroborated by a deviation of the $B_{hf}$ from molecular field model curve with a decrease of $\Delta E_Q$ below Néel temperature. This may be ascribed to contribution of orbital angular momentum below 27 K, above which gets quenched by crystal field of the Fe$^{2+}$ ion in LiFePO$_4$ [31, 22].

4.  **Summary**

DC magnetization and low-temperature Mössbauer spectroscopy measurements on LiFePO$_4$ have established the antiferromagnetic ordering in this compound which begins at 50 K. These measurements have also revealed the $Fe^{2+}$ ion $(3d^6, {}^5D)$ in LiFePO$_4$ undergo relaxation phenomena of the spin-lattice type. This is evidenced by the asymmetric broadening of the Mössbauer absorption peaks. The temperature dependence of the quadrupole splitting indicates the excited orbital states mix strongly to the orbital doublet ground state via spin-orbit coupling at low temperatures. The anomaly in magnetization and Mössbauer parameters below 27 K may be ascribed to the contribution of orbital angular momentum, above which gets quenched by crystal field Extended measurements of Mossbauer spectra at sub-helium temperatures as well as neutron diffraction measurement would probably give a more definitive quantitative picture of this fascinating phenomenon.

# Figure Captions

Figure 1(a). The X-ray diffraction pattern of LiFePO$_4$ at 300K. The inset shows the crystal structure of LiFePO$_4$ at 300 K obtained from the Rietveld refinement for atomic positions.

Figure 1(b). FTIR spectrum of LiFePO$_4$ recorded at 300 K. The inset shows the FESEM image of LiFePO$_4$.

Figure 2. Field-cooled magnetization plot of LiFePO$_4$ with temperature. The inset shows the reciprocal susceptibility below T$_N$ (~50 K).

Figure 3. Mössbauer spectra of LiFePO$_4$ in the (a) paramagnetic and (b) antiferromagnetic regime. The red line indicates the fit through the data.

Figure 4. Temperature dependence of (a) quadrupole splitting $\Delta E_Q$ and (b) hyperfine field B$_{hf}$. Figures (c) and (d) show the temperature dependence of relaxation frequency $\left(\tau^{-1}\right)$ below and above Néel temperature T$_N$.

**Figure 1(a)**

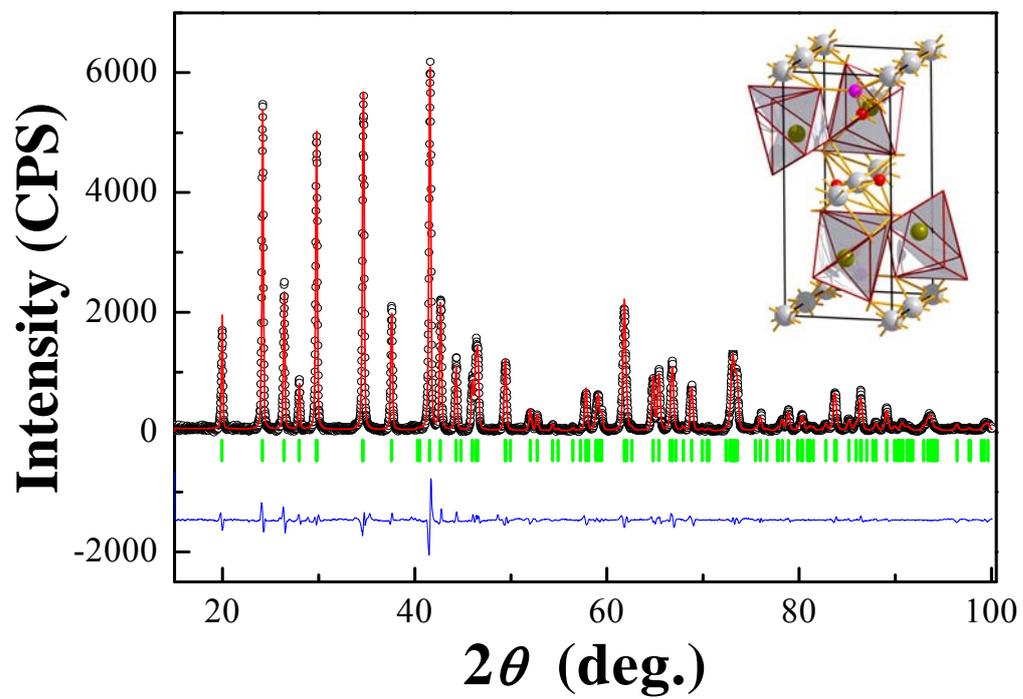

**Figure 1(b)**

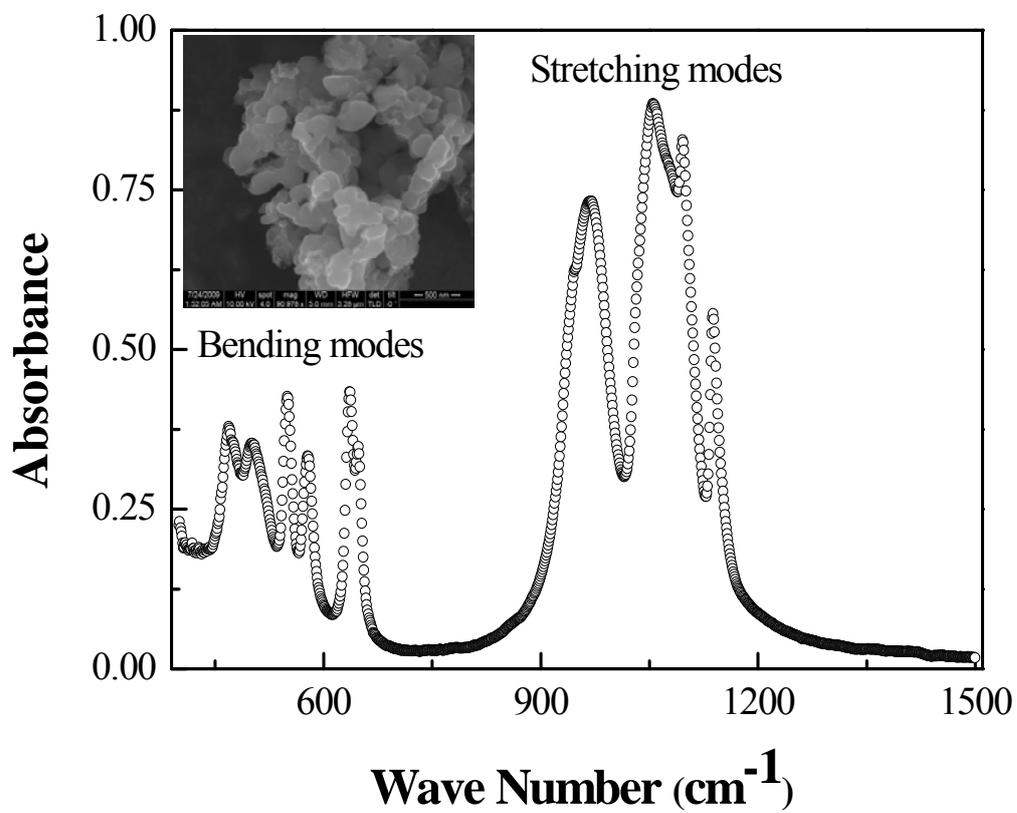

**Figure 2**

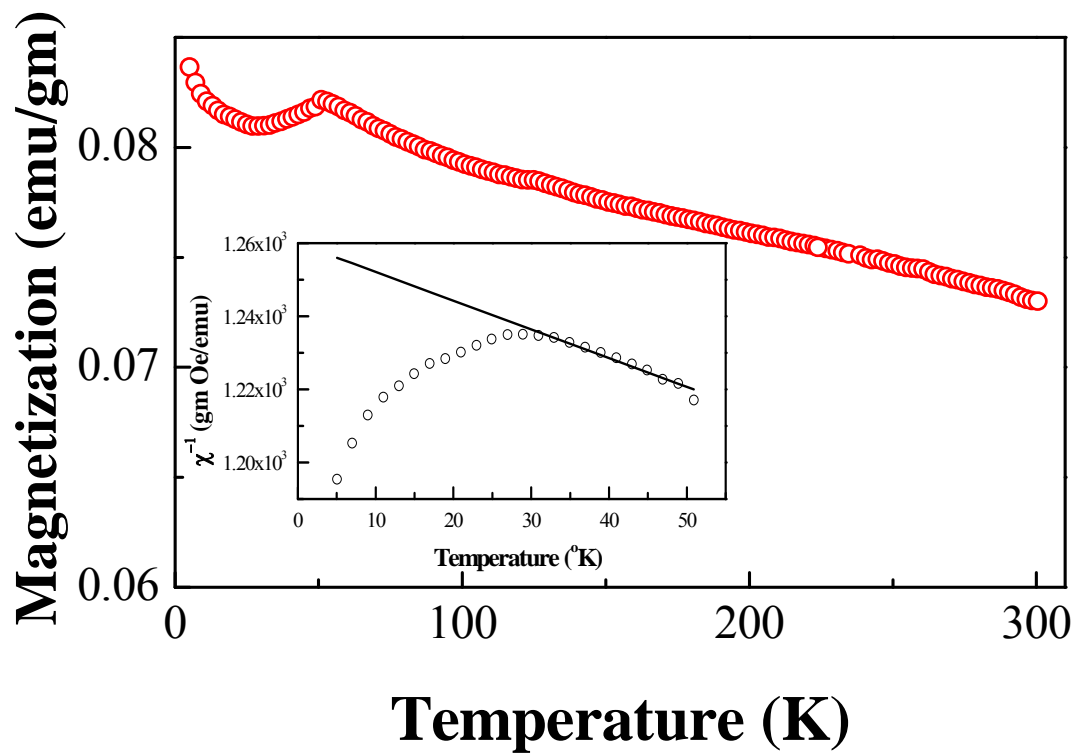

**Figure 3(a)**

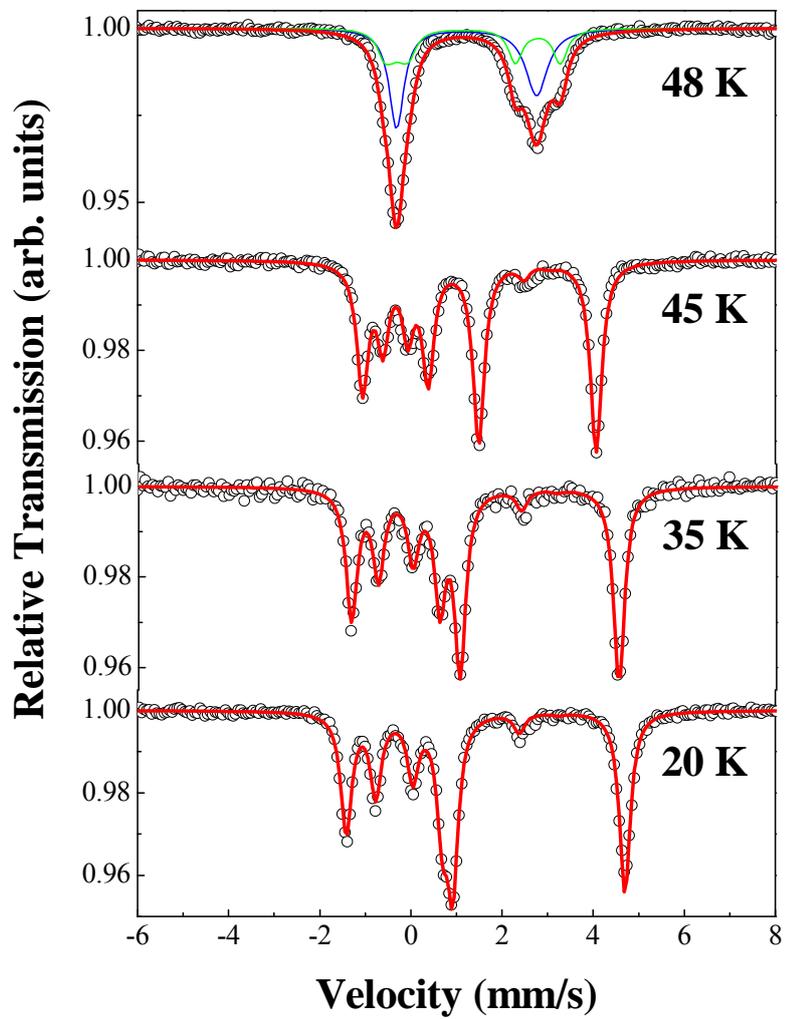

**Figure 3(b)**

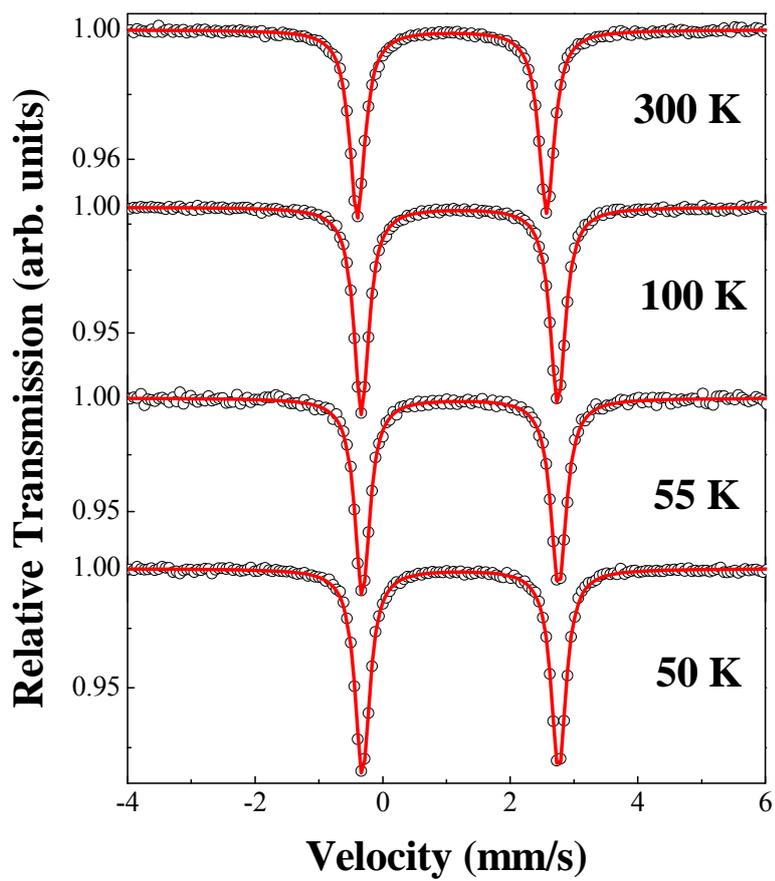

**Figure 4**

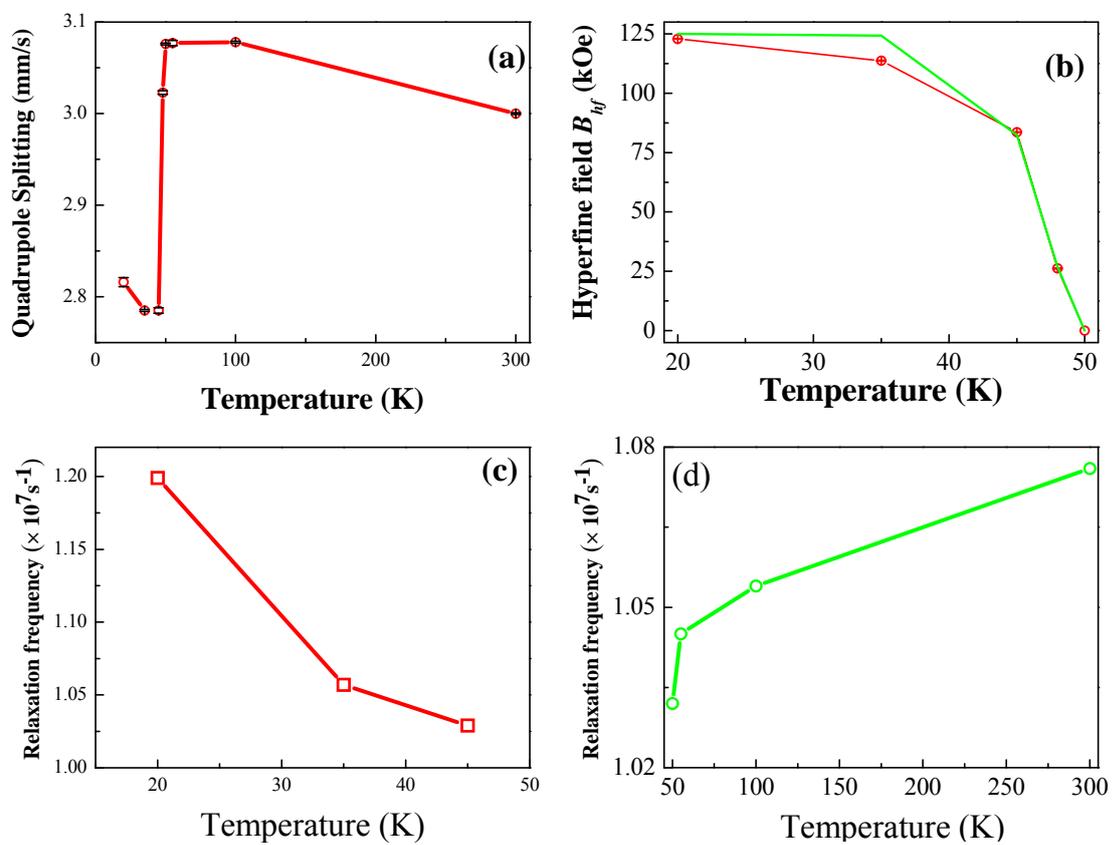

# Tables

TABLE I. Atomic positions within *Pnma* of LiFePO$_4$ at 300 K determined by Rietveld analysis. Lattice parameters are *a* = 10.34263 Å, *b* = 6.01785 Å and *c* = 4.70136 Å.

| Atom | Site | x | y | z | B$_{iso}$ |
|------|------|---------|---------|---------|----------|
| Li | 4*a* | 0.00000 | 0.00000 | 0.00000 | 0.05828 |
| Fe | 4*c* | 0.21751 | 0.25000 | 0.52632 | -0.00430 |
| P  | 4*c* | 0.40404 | 0.25000 | 0.07954 | -0.02624 |
| O1 | 4*c* | 0.40653 | 0.25000 | 0.75295 | 1.00000 |
| O2 | 4*c* | 0.04419 | 0.25000 | 0.29682 | 1.00000 |
| O3 | 8*d* | 0.33448 | 0.04929 | 0.21717 | 1.00000 |

TABLE II. Mössbauer parameters of LiFePO$_4$ in paramagnetic regime:

| Sl. No. | Temperature (K) | IS (mm/s) | QS (mm/s) | W (mm/s) | A$_{21}$ (mm/s) | W$_{21}$ (mm/s) |
|---|---|---|---|---|---|---|
| 1 | 300 | 1.286 (1) | 2.968 (1) | 0.292 (2) | 0.973 | 0.998 |
| 2 | 100 | 1.306 (1) | 3.078 (1) | 0.281 (2) | 0.971 | 1.023 |
| 3 | 55 | 1.320 (1) | 3.077 (3) | 0.282 (4) | 0.983 | 1.016 |
| 4 | 50 | 1.320 (2) | 3.076 (1) | 0.281 (2) | 0.978 | 1.003 |
| 5 | 48 | 1.316 (2) | 3.017 (2) | 0.350 (3) | 0.986 | 1.463 |

Here IS − Isomer shift, QS − Quadrupole splitting, W − Full width at half maximum (FWHM), A$_{21}$ − Ratio of the areas of higher to lower velocity peaks and W$_{21}$ − Ratio of the FWHMs of higher to lower velocity peaks.

TABLE III. Mossbauer parameters of LiFePO$_4$ in Antiferromagnetic regime:

| Sl. No. | Temperature (K) | IS (mm/s) | QS (mm/s) | W (mm/s) | B$_{hf}$ (kOe) | $\eta$ |
|---|---|---|---|---|---|---|
| 1 | 20 | 1.322 (1) | 2.816 (5) | 0.292 (2) | 122.9 (1) | 0.762 (6) |
| 2 | 35 | 1.323 (1) | 2.785 (1) | 0.294 (2) | 113.7 (2) | 0.805 (2) |
| 3 | 45 | 1.319 (1) | 2.785 (8) | 0.306 (5) | 83.6 (1) | 0.812 (6) |
| 4 | 48 | 1.316 (2) | 3.023 (2) | 0.437 (8) | 26.2 (2) | 0.850 (7) |

Here B$_{hf}$ – the hyperfine field and $\eta$ – the asymmetry parameter